\documentclass{aa}  

\usepackage{natbib}
\usepackage{graphicx}
\usepackage{txfonts}
\usepackage[breaklinks]{hyperref}  
\usepackage{breakurl}              
\usepackage{color}
\usepackage{url,twoopt,amssymb}
\usepackage{multirow}

\hypersetup{
  colorlinks=true,   
  urlcolor=blue,     
  linkcolor=blue     
}

\begin{document}
  \title{Evaluation of the ICRF stability from position time series analysis}
  \author{N. Liu\inst{1,2}
     \and
          S. Lambert\inst{3} 
          \and
          F. Arias\inst{3}
          \and
          J.-C. Liu\inst{1}
          \and
          Z. Zhu\inst{1}
          }

  \institute{School of Astronomy and Space Science,
   		     Key Laboratory of Modern Astronomy and Astrophysics (Ministry of Education), Nanjing University, Nanjing 210023, P. R. China\\
              \email{[niu.liu;zhuzi]@nju.edu.cn}
         \and
             School of Earth Sciences and Engineering, Nanjing University, Nanjing 210023, P. R. China
         \and
             SYRTE, Observatoire de Paris,
             Universit\'e PSL, CNRS, Sorbonne Universit\'e, LNE, Paris, France\\
             \email{sebastien.lambert@obspm.fr}
             }

  \date{Received; accepted}

  \abstract
  {
  The celestial reference frame is realized by absolute positions of extragalactic sources that are assumed to be fixed in the space.
  The fixing of the axes is one of the crucial points for the International Celestial Reference System (ICRS) concept.
  However, due to various effects such as its intrinsic activity, the apparent position of the extragalactic sources may vary with time, resulting in a time-dependent deviation of the frame axes that are defined by the positions of these sources.
  }
  {
  We aim to evaluate the axis stability of the third realization of the International Celestial Reference Frame (ICRF3).
  }
  {
  We first derive the extragalactic source position time series from observations of very long baseline interferometry (VLBI) at the dual $S/X$-band (2.3/8.4~GHz) between August 1979 and December 2020.
  We measure the stability of the ICRF3 axes in terms of the drift and scatter around the mean:
  (i) we estimate the global spin of the ICRF3 axes based on the apparent proper motion (slope of the position time series) of the ICRF3 defining sources;
  (ii) we also construct the yearly representations of the ICRF3 through annually averaged positions of the ICRF3 defining sources and estimate the dispersion in the axis orientation of these yearly frames.
  }
  {
  The global spin is no higher than $\mathrm{0.8\,\mu as\,yr^{-1}}$ for each ICRF3 axis with an uncertainty of $\mathrm{0.3\,\mu as\,yr^{-1}}$,
  corresponding to an accumulated deformation smaller than $\mathrm{30\,\mu as}$ for the celestial frame axes during 1979.6--2021.0.
  The axes orientation of the yearly celestial frame becomes more stable as time elapses, with a standard deviation of 10--20$\mathrm{\,\mu as}$ for each axis.
  }
  {
  The axes of the ICRF3 are stable at approximately 10--20~$\mathrm{\mu as}$ from 1979.6--2021.0 and the axis stability does not degrade after the adoption of the ICRF3.
  }


\maketitle


\section{Introduction}

    The celestial reference frame provides the basic position reference for astronomy and geosciences.
    The current realization of the celestial reference frame as adopted by the International Astronomical Union (IAU) in 2018 is the third realization of the International Celestial Reference Frame \citep[ICRF3;][]{2020A&A...644A.159C}.
    The ICRF3 is constructed by positions of more than 4000 extragalactic sources based on very long baseline interferometry (VLBI) observations made at the dual $S/X$-, $K$-, and dual $X/Ka$- bands (2/8~GHz, 24~GHz, and 8/32~GHz) since 1979.
    With several improvements in the modeling, for example, the correction of the Galactic aberration effect due to the acceleration of the solar system barycenter \citep{2019A&A...630A..93M}, and more accumulated data, the best position precision achieved for individual sources in the ICRF3 catalog has reached a level of 30 microarcseconds ($\mathrm{\mu as}$).
    
    The extragalatic sources are used to construct the celestial reference frame because they are distant from us so that they appear more compact and stable (i.e., showing a negligible motion) than other objects such as Galatic stars.
    However, the apparent positions of the extragalatic sources vary with time, both in the radio and optical domain \citep[e.g., see][]{2009jsrs.conf..199A,2013A&A...553A.122L}.
    This kind of position variability of extragalatic sources, also referred to as the astrometric instability, would lead to variations in the direction of the celestial frame axes defined by positions of those sources.
    This phenomenon is known as the celestial frame instability \citep[e.g., see][]{2008A&A...481..535L}, a common issue for all kinds of extragalatic celestial reference frames such as the ICRF3 and the \textit{Gaia} celestial frame \citep[\textit{Gaia}-CRF;][]{2018A&A...616A..14G}.
    The VLBI celestial frame instability would introduce additional noise to the VLBI products such as the Earth orientation parameters (EOPs); the implication would be more serious for the nutation series since thus so far the VLBI is the sole technique that can measure the nutation angle.
    \citet{2003JGRB..108.2275D} and \citet{2008A&A...481..535L} study the impact of the celestial frame instability on the estimate of the nutation terms from the VLBI observations and report an additional noise of 15 microarcseconds ($\mathrm{\mu as}$) in the amplitude of the 18.6 yr nutation term.
    The variability of extragalatic sources may also degrade the precision of the VLBI-\textit{Gaia} frame alignment \citep{2013A&A...552A..98T,2016A&A...587A.112T,2018A&A...611A..52T,2018MNRAS.474.4477L}.
    Therefore, it is necessary to assess and monitor the instability level of the celestial reference system, which is the main motivation of this work.
    
    There are several possibilities that make radio sources apparently unstable.
    The main origin is the temporal evolution of the radio source structure due to intrinsic variability such as the ejection of new jet features, which could manifest itself as an apparent proper motion \citep{2011AJ....141...91F,2011AJ....141..178M} or positional offset along the jet \citep{2016MNRAS.455..343P}.
    Other causes include the difference in radio source position as ``seen'' by different VLBI networks \citep{2003JGRB..108.2275D} and the weak microlensing effect \citep{1998MNRAS.300..287S,2017ApJ...835...51L,2020ApJ...898...51L}.    
    
    Several authors propose indicators to characterize the source astrometric stability.
    Statistics based on the source coordinate time series are often used, for example, slope and standard deviation \citep{2003A&A...403..105F}, Allan variance \citep{2018A&A...618A..80G}.
    The ICRF Working Group uses a quantity that considers the coordinate variations (weighted root-mean-square, WRMS) and the reduced chi-square for right ascension and declination \citep{2015AJ....150...58F,2020A&A...644A.159C}.
    In addition, other authors construct indicators based on the interstellar scintillation around the source \citep{2013MNRAS.434..585S} and flux density time series \citep[light curve;][]{2014JGeod..88..575S}, which are found to correlate with the astrometric stability.
    Based on these indicators, sources with an unstable position can be recognized and ruled out, leaving only stable sources to define and maintain the axes of the celestial reference frame, which are known as the defining sources \citep{2003A&A...403..105F,2006A&A...452.1107F,2009A&A...493..317L}.
    The stability of the VLBI celestial reference frame can thus be improved when a proper ensemble of stable sources is used as the defining source subset \citep{2004A&A...422.1105A,2017MNRAS.466.1567L}.
    A list of 303 sources with good astrometric behavior known as the ICRF3 defining sources was carefully selected, whose positions implicitly define the axes of the ICRF3 \citep{2020A&A...644A.159C}.
    
    The ICRF axes were found to be stable on a level of $\mathrm{20\,\mu as}$ \citep{1998AJ....116..516M}, value of which has been further improved to $\mathrm{10\,\mu as}$ for the ICRF2 \citep{2015AJ....150...58F}.
    These results are based on comparing the relative orientation of various subsets of sources.
    Adopting a different method, \citet{2013A&A...553A.122L} finds that the axis stability of the ICRF2 does not degrade after 2009, which is approximately $\mathrm{20\,\mu as}$ for each axis.
    Later, \citet{2014A&A...570A.108L} compares VLBI radio source catalogs submitted from different analysis centers of the International VLBI Service for Geodesy and Astrometry \citep[IVS;][]{2017JGeod..91..711N} and reports an agreement of several tens of $\mathrm{\mu as}$ among these catalogs.
    Recently, \citet{2018A&A...618A..80G} study the astrometric stability of extragalactic sources in the light of the Allan standard deviation of VLBI position time series.
    They conclude that the source position showing a stable behavior is likely to become unstable within a longer time span.
    This highlights the need to regularly monitor the astrometric behavior of extragalactic sources and the axis stability of the VLBI celestial reference frame, as already pointed out in \citet{2013A&A...553A.122L}.
    
    This work aims to assess the stability of ICRF3 axes based on the extragalactic source coordinate time series data, which have the advantage of being independent from VLBI global catalogs.
    Two methods were proposed for this purpose.
    One is to consider the accumulated effect due to the global spin of the VLBI celestial frame estimated from the apparent proper motion of extragalatic sources.
    The other is to evaluate the variation of the axis orientation for yearly representations of the ICRF3.
    

\section{Data and their preparation}  \label{sec:data}

    The extragalactic source coordinate time series were derived from ten separate global solutions based on the same VLBI observations spanning from August 1979 to December 2020. 
    The data consisted of approximately 16.3~M group delays made at dual $S/X$-band (2.4/8.3~GHz), in total 6974 sessions, which are publicly available at the IVS website\footnote{
    The data can be found at \url{https://ivscc.gsfc.nasa.gov/products-data/data.html}.
    A regular session is a collection of VLBI observations made within 24 hours. 
    The full session list can be found at \href{https://doi.org/10.5281/zenodo.5795064}{DOI:10.5281/zenodo.5795064}.}.
    We used the geodetic VLBI analysis software package Calc/Solve \citep[version 20200123;][]{1986AJ.....92.1020M} to process these VLBI observations.
    The ten VLBI global solutions followed the same setup and parameterization, except some special configurations of radio source position estimate and celestial frame maintenance as explained in the following.
    Usually, the source position is assumed to not change with time and is treated as a global parameter in the regular VLBI data analysis, which means that only one measurement of position can be obtained for each source.
    A not-net-rotation (NNR) constraint is applied to the positions of the defining sources in order to fix the orientation of the celestial frame.
    To obtain source position time series, the position of one tenth of the sources (including one tenth of the defining sources) was downgraded as the sessionwise parameter in each of our solutions.
    These ten groups of sources were chosen so that they were distributed as uniformly as possible over the sky.
    The orientation of the resulting celestial frame is fixed by applying the NNR constraint to the remaining nine tenths of the defining sources.
    As a result, we obtained the coordinate time series for every tenth of the sources from each VLBI solution.
    This method of generating the VLBI coordinate time series was first proposed in \citet{2003A&A...403..105F}, for  this study we developed an improved version.
    The general technique description for these solutions can be found at the Paris Observatory Geodetic VLBI Center\footnote{See \url{http://ivsopar.obspm.fr/24h/opa2021a.eops.txt}.};
    for more information about the special configurations, we refer to \citet{2013A&A...553A.122L} and \citet{2021A&A...648A.125G}.
    The root-mean-square of the postfit residual delay is approximately 40~picoseconds and the reduced $\chi^2$ is 2.7.
    
    A common step after the VLBI solution is to align the result reference frame onto the ICRS by a rigid rotation.
    We noted that there is a global rotation of a few tens of $\mathrm{\mu as}$ between the ICRF3 and the opa2019a solution as reported in the IERS annual report\footnote{\url{https://hpiers.obspm.fr/icrs-pc/newwww/analysis/icrsra_2019_VLBI.pdf}}, the latter being constructed with almost the same analysis strategies as our solutions.
    We found a rotation of $\vec{R_0}=(+36,-52,-10)^{\rm T}\,\mu{\rm as}$ of the latest opa2021a solution\footnote{\url{http://ivsopar.obspm.fr/24h/opa2021a.crf}.} with respect to the ICRF3 using the sample of all the ICRF3 defining sources and the method described later in this section.
    We decided to apply a global rotation of $-\vec{R_0}$ to our ten solutions so that there would be no bias in the orientation between our solutions and the ICRF3 in a mean sense.
    Since we concentrate on the time-dependent stability of the celestial frame, such a constant rotation will not alter our results.
    
    We obtained the coordinate time series for 5290 extragalactic sources, among which 4554 sources are common to the ICRF3 catalogs, including all 4536 sources in the $S/X$-band catalog and all 303 ICRF3 defining sources.
    The median number of sessions in which a given source was observed was 5 for all sources and it increased to 139 for the ICRF3 defining sources.
    The median value of the mean observation epoch for individual sources is 2015.15 for the whole sample and 2013.11 for the ICRF3 defining source subset.
    The typical uncertainty (median value) in the coordinates for a given source is approximately 0.2\,mas for the right ascension and 0.3\,mas for the declination. 
    
    The axis stability of the celestial frames was evaluated in terms of the long-term drift and the wandering around the mean, as remarked below.\\
    1. We estimated the apparent proper motions (slope) from coordinate time series and then fitted the global spin of the celestial reference frame based on these apparent proper motions.
       The spin multiplied by the time span of the VLBI observations gives an estimate of the axis stability.\\
    2. We constructed the yearly representations of the ICRF3 from the coordinate time series and compared the relative orientation of these yearly celestial reference frames referred to the ICRF3 $S/X$-band catalog.
       The dispersion of the relative orientation angles provides another assessment of the axis stability.
    
    The global (large-scale) features in any vector field on the celestial sphere can be described by vector spherical harmonics \citep[VSH;][]{2012A&A...547A..59M}.
    Here we only considered the first degree of the VSH, which consists of a rotation vector $\vec{R}=(R_1,R_2,R_3)^{\rm T}$ and a glide vector $\vec{G}=(G_1,G_2,G_3)^{\rm T}$.
    The full equation can be expressed as
    \begin{equation} \label{eq:vsh01}
        \begin{array}{ll}
            \Delta_{\alpha^*}  = &-R_1\cos\alpha\sin\delta  - R_2\sin\alpha\sin\delta + R_3\cos\delta \\
                        &-G_1\sin\alpha            + G_2\cos\alpha, \\
            \Delta_{\delta}    = &+R_1\sin\alpha            - R_2\cos\alpha \\
                        &-G_1\cos\alpha\sin\delta  - G_2\sin\alpha\sin\delta + G_3\cos\delta. \\
        \end{array}
    \end{equation}
    The notation $\Delta_{\alpha^*}=\Delta_{\alpha}\cos\delta$ will be used throughout this paper.
    The rotation vector consists of rotation angles around the X-, Y-, and Z-axis and is used to characterize the stability of the ICRF3 axes.
    
    Since we dealt with the rotation vectors both from the position offset and apparent proper motion field, different notations were used for clarity.
    Following conventions, for example, in \citet[][]{2018A&A...616A..14G}, we used the orientation offset vector $\vec{\epsilon}=(\epsilon_x,\epsilon_y,\epsilon_z)^{\rm T}$ to represent the rotation vector estimated from the position offset, meaning that ($\Delta_{\alpha^*}$, $\Delta_{\delta}$) in Eq.~(\ref{eq:vsh01}) were substituted by ($\Delta\alpha\cos\delta$, $\Delta\delta$).
    We denoted the rotation vector estimated from the apparent proper motion as $\vec{\omega}=(\omega_x,\omega_y,\omega_z)^{\rm T}$.
    This vector models the global spin of the celestial frame, that is, the change rate of the ICRF3 axes directions.
    
    The rotation signal was estimated together with the glide parameters in the least-square fitting.
    The data were weighted by the full covariance matrix between right ascension and declination (i.e., including the covariance between the right ascension and declination).
    No outlier elimination was implemented in order to avoid the reduction of the rotation estimate caused by removing any source with a significant position offset or apparent proper motion in the elimination progress.


\section{Analysis and results}  \label{sec:results}


\subsection{Global spin from apparent proper motion}  \label{subsec:spin-from-apm}

    \begin{figure}
        \centering
        \includegraphics[width=\columnwidth]{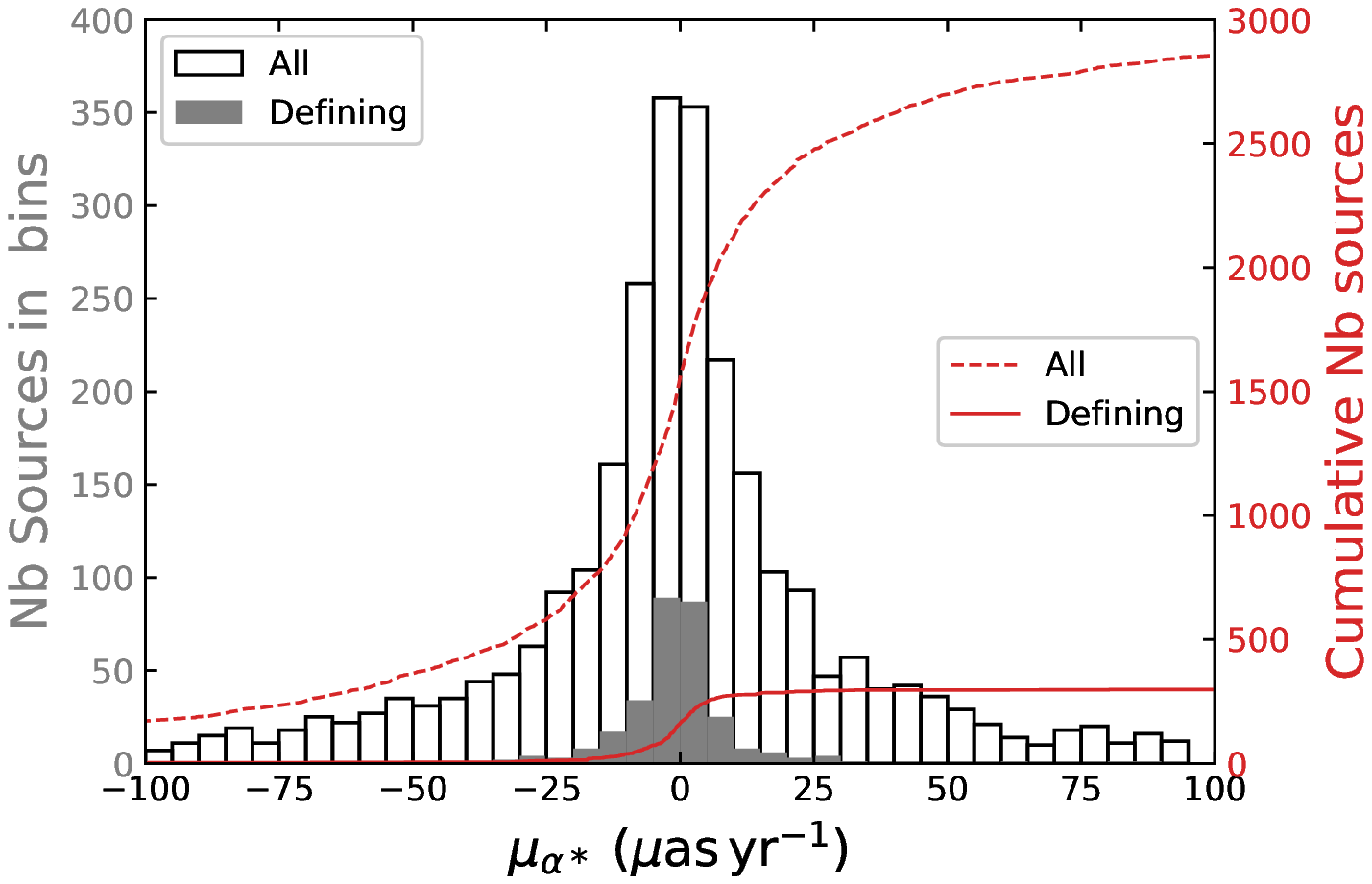}
        \includegraphics[width=\columnwidth]{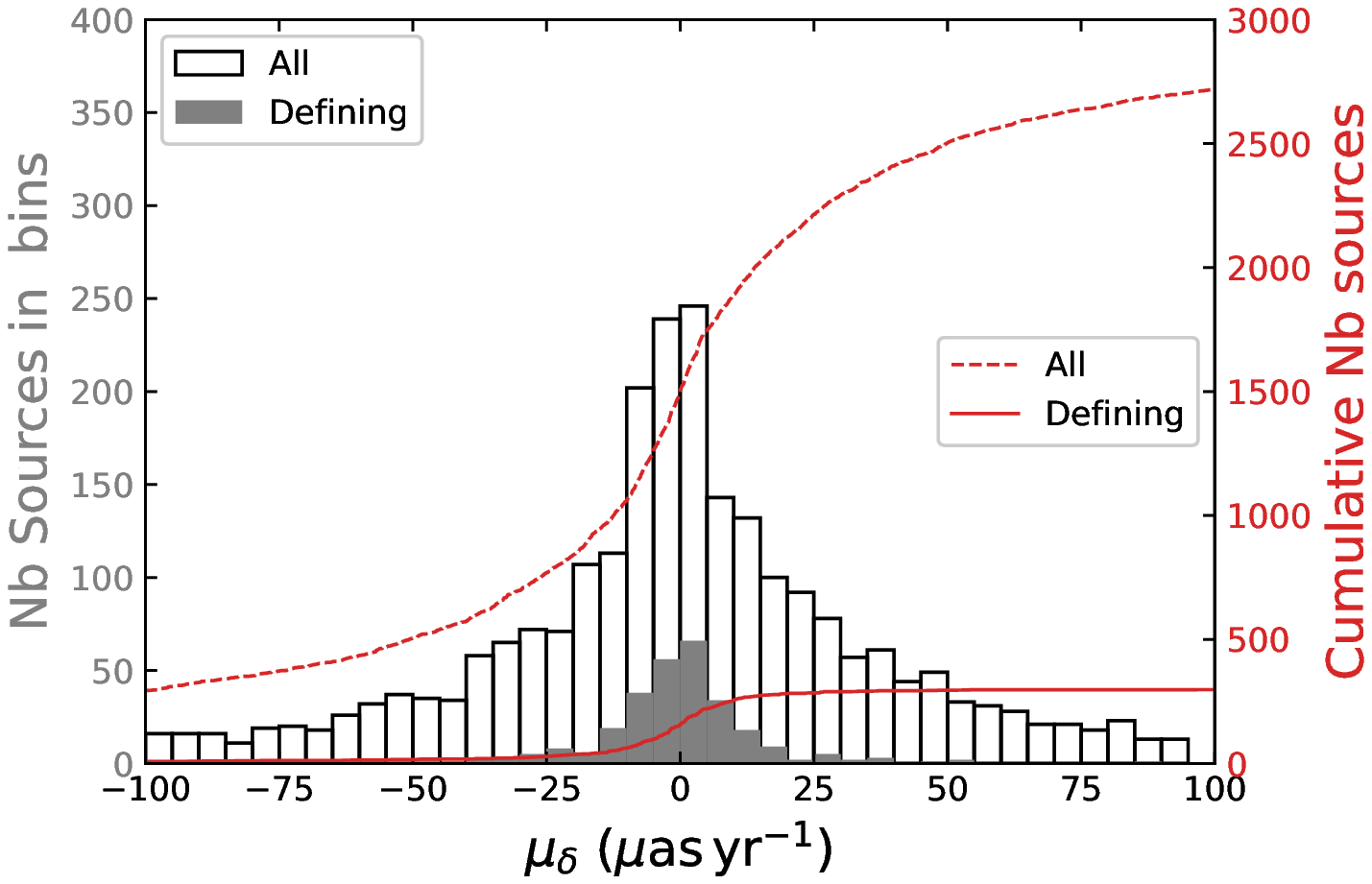}
        \caption{\label{fig:apm-dist}%
            Distribution of the apparent proper motion in the right ascension ($upper$) and declination ($lower$) for 3034 extragalctic sources fitted from their coordinate time series.
            The left and right vertical axes indicate the number of sources in the each bin and cumulative from the leftmost bin.
            The distribution for 299 sources among the so-called 303 ICRF3 defining source list is labeled with grey and red lines.}
    \end{figure}

    \begin{figure}
        \centering
        \includegraphics[width=\columnwidth]{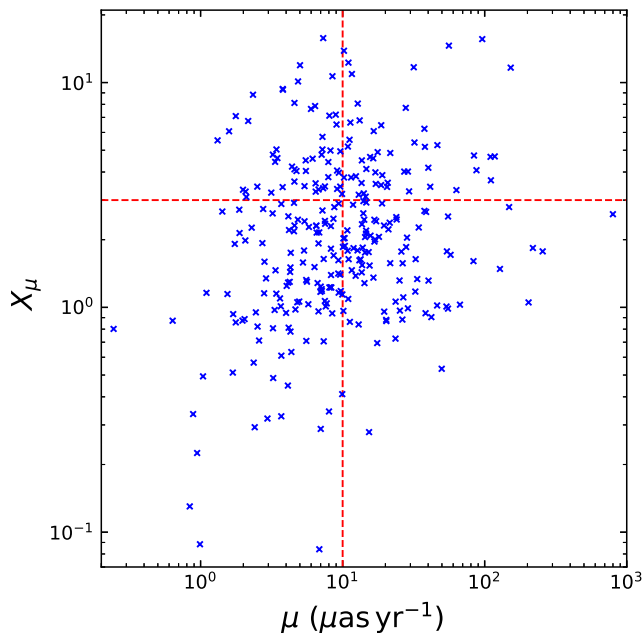}
        \caption{\label{fig:apm-sig}%
            Distribution of the total apparent proper motion $\mu$ and its significance $X_\mu$ for the 299 sources among the ICRF3 defining
            sources.
            Sources located in the top right corner are considered as having a large and also significant apparent proper motion, i.e., $\mu\,>\,10\,\mathrm{\mu as\,yr^{-1}}$ and $X_\mu\,>\,3$.
            }
    \end{figure}

    \begin{figure}
        \centering
        \includegraphics[width=\columnwidth]{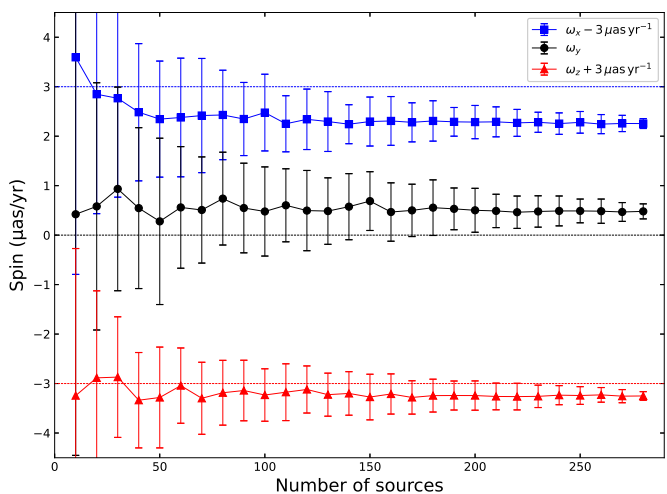}
        \caption{\label{fig:spin-dist}%
        Global spin estimated from the apparent proper motion versus the sample size used in the bootstrap sampling.
        The markers and errorbars stand for the mean value and standard deviation from 100 bootstrap samples, respectively.
        }
    \end{figure}
    
    We estimated the apparent proper motions of the radio sources based on their coordinate time series through a weighted least squares fitting.
    This model is described as follows.
    \begin{equation} \label{eq:apm}
       \alpha = \mu_{\alpha} (t-t_0) + \alpha_0,\ \delta = \mu_\delta (t-t_0) + \delta_0, 
    \end{equation}
    where $\mu_{\alpha*}=\mu_{\alpha}\cos\delta$ and $\mu_\delta$ are the apparent proper motions in right ascension and declination, respectively, and $t_0$ is the mean epoch of the observation span for a given source.
    
    All the data points were weighted by the inverse of the full covariance matrix for the right ascension and declination, which means that $\mu_{\alpha*}$ and $\mu_\delta$ were fitted simultaneously.
    The data points whose distance to the mean position for a given source was greater than three times their formal errors in either right ascension or declination were considered as outliers and thus removed before estimating $\mu_{\alpha*}$ and $\mu_\delta$.
    Sources with less than five data points in the remaining time series were also removed from our list for the sake of a reliable proper motion determination.
    By doing so, we obtained the apparent proper motion for 3034 sources, 299 belonging to the ICRF3 defining source subset.
    
    Figure~\ref{fig:apm-dist} depicts the distribution of the fitted proper motion.
    The median value of the apparent proper motion for all the sources is $\mathrm{-0.36~\mu as\,yr^{-1}}$ in the right ascension and $\mathrm{0.26~\mu as\,yr^{-1}}$ in the declination; they are $\mathrm{-0.57~\mu as\,yr^{-1}}$ and $\mathrm{-0.33~\mu as\,yr^{-1}}$ for the ICRF3 defining source subset.
    For more than half of the sources in the whole sample, the apparent proper motion is within $\mathrm{\pm\,30~\mu as\,yr^{-1}}$ both in right ascension and declination.
    
    We were more concerned about the distribution of the apparent proper motions for the defining sources.
    We computed the total apparent proper motion as 
    \begin{equation} \label{eq:tot-apm}
       \mu = \sqrt{\mu_{\alpha*}^2 + \mu_\delta^2}.
    \end{equation}
    Approximately 55\% of the defining sources in our sample yield an apparent proper motion of less than $10\,\mathrm{\mu as\,yr^{-1}}$.
    To quantify the significance of the apparent proper motion, we also computed a normalized quantity $X_\mu$ as
    \begin{equation} \label{eq:nor-apm}
       X_\mu^{2}=\left[\begin{array}{ll}
        \mu_{\alpha*} & \mu_{\delta}
        \end{array}\right]\left[\begin{array}{cc}
        \sigma_{\mu_{\alpha*}} & \rho_{\mu_{\alpha*},\mu_{\delta}} \sigma_{\mu_{\alpha*}} \sigma_{\mu_{\delta}} \\
        \rho_{\mu_{\alpha*},\mu_{\delta}} \sigma_{\mu_{\alpha*}} \sigma_{\mu_{\delta}} & \sigma_{\mu_{\delta}}
        \end{array}\right]^{-1}\left[\begin{array}{l}
        \mu_{\alpha*} \\
        \mu_{\delta}
        \end{array}\right],
    \end{equation}    
    as proposed in \citet{2016A&A...595A...5M}.
    We found that approximately 16\% of sources showed both $\mu\,>\,10\,\mathrm{\mu as\,yr^{-1}}$ and $X_\mu > 3$, that is, large and also statistically significant apparent proper motions.
    
    We fitted the global spin from the apparent proper motion based on the 299 sources among the ICRF3 defining source list and reported the result below.
    \begin{equation} \label{eq:spin-from-def}
         \begin{array}{l}
             \omega_x = -0.73\,\mathrm{\mu as\,yr^{-1}} \pm 0.27 \,\mathrm{\mu as\,yr^{-1}}, \\
             \omega_y = +0.48\,\mathrm{\mu as\,yr^{-1}} \pm 0.31 \,\mathrm{\mu as\,yr^{-1}}, \\
             \omega_z = -0.25\,\mathrm{\mu as\,yr^{-1}} \pm 0.21 \,\mathrm{\mu as\,yr^{-1}}. 
         \end{array}
    \end{equation}
    The total spin was estimated to be $\omega = 0.91\,\mathrm{\mu as\,yr^{-1}} \pm 0.28\,\mathrm{\mu as\,yr^{-1}}$, pointing in the direction of ($\alpha\,=\,147\,^{\circ}\,\pm\,20,^{\circ}$, $\delta\,=\,-16\,^{\circ}\,\pm\,14,^{\circ}$).
    
    We noted that the estimate from the least squares fitting might be biased by unreliable apparent proper motions of individual sources, and we also noted the fact that the uncertainty given by the least squares fitting is always underestimated.
    We used the bootstrap sampling technique to generate 1000 new samples (i.e., select 299 sources with replacement) and adopted the mean and standard deviation of the spin parameters derived from these samples as a robust estimate of the unknowns and the associated formal uncertainty.
    The results are tabulated below.
    \begin{equation} \label{eq:spin-from-def-from-bs}
         \begin{array}{l}
             \omega_x = -0.71\,\mathrm{\mu as\,yr^{-1}} \pm 0.42 \,\mathrm{\mu as\,yr^{-1}}, \\
             \omega_y = +0.51\,\mathrm{\mu as\,yr^{-1}} \pm 0.57 \,\mathrm{\mu as\,yr^{-1}}, \\
             \omega_z = -0.24\,\mathrm{\mu as\,yr^{-1}} \pm 0.36 \,\mathrm{\mu as\,yr^{-1}}. 
         \end{array}
    \end{equation}
    One can easily find that the estimates in the Eqs.~(\ref{eq:spin-from-def}) and (\ref{eq:spin-from-def-from-bs}) are consistent, but the bootstrap errors should be more realistic.
    
    We wanted to know whether the axes stability of the celestial frame degrades when not all the ICRF3 defining sources are observed, as is the usual case for the VLBI campaigns.
    We randomly picked $N$ sources from the sample (without replacement) and computed the rotation parameters.
    This procedure was repeated 100 times and we calculated the mean and standard deviation of the rotation parameters.
    The sample size $N$ varies from 10 to 290, with a step of 10.
    Figure~\ref{fig:spin-dist} presents the distribution of the mean spin (marker) and the standard deviation (errorbar) as a function of $N$.
    We found that the spin parameters were generally stable when $N \gtrsim 50$, where $\mathrm{\omega_x\sim-0.60\,\mu as\,yr^{-1}}$,  $\mathrm{\omega_y\sim+0.53\,\mu as\,yr^{-1}}$, and  $\mathrm{\omega_z\sim-0.24\,\mu as\,yr^{-1}}$.
    The smaller errorbar at large values of $N$ suggests that the estimate of the global spin converges as expected when there is a sufficiently large number of the ICRF3 defining sources, regardless of which sources are included.
    This experiment suggests that at least 50 defining sources should be included in the VLBI observation schedule and data analysis so that the output of the VLBI solution will be less influenced by the issues related to the frame orientation.
        
    We also repeated the procedures for all the sources with an estimate of the apparent proper motion.
    The results agree well with those given in Eqs.~(\ref{eq:spin-from-def})-(\ref{eq:spin-from-def-from-bs}) within the standard deviation.
    
    The studies above suggest that only the spin around the X-axis is slightly greater than the formal uncertainty from zero (confident at 1.7-$\sigma$, c.f.~Eq.~(\ref{eq:spin-from-def-from-bs})), therefore the spin in each axis of the ICRF3 cannot be considered as significant.
    Considering a time span of 41.4\,yr long (1979.6--2021.0) and the subset of ICRF3 defining sources, the accumulated deformation is about $\mathrm{30\,\mu as}$, $\mathrm{20\,\mu as}$, and $\mathrm{10\,\mu as}$ for the X-, Y-, and Z- axes of the celestial reference frame, respectively.
    The uncertainty is approximately $\mathrm{20\,\mu as}$ considering the bootstrap errors (Eq.~(\ref{eq:spin-from-def-from-bs})) and $\mathrm{10\,\mu as}$ using the least-square errors (Eq.~(\ref{eq:spin-from-def})).
    As a result, a conservative estimation of the ICRF3 axis stability for this method is approximately 10$\mathrm{\,\mu as}$--20$\mathrm{\,\mu as}$.


\subsection{Orientation stability of the yearly celestial reference frames}  \label{subsec:orient-from-yearly-crf}

    \begin{figure}
        \centering
        \includegraphics[width=\columnwidth]{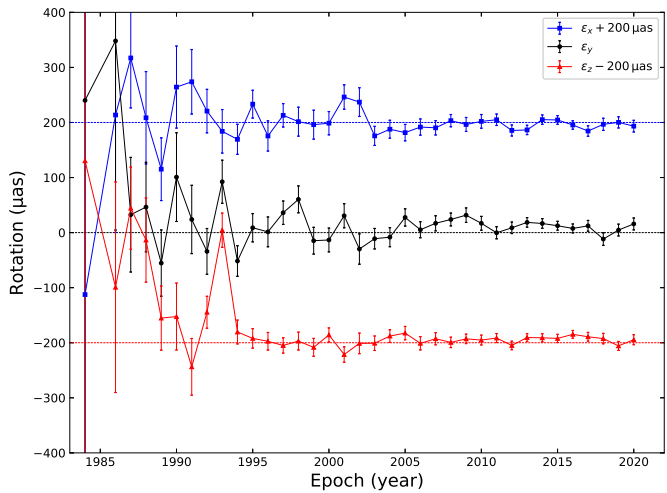}
        \caption{\label{fig:rot-yearly-ts}   
         Relative orientation of yearly celestial reference frames with respect to the ICRF3 $S/X$-band frame. 
         The yearly celestial frames were constructed from the annual mean positions of the ICRF3 defining sources.
         The marker indicates the estimate of the orientation parameters, while the bar shows the formal uncertainty from the least-squares fitting.
         }
    \end{figure}
    
    \begin{figure}
        \centering
        \includegraphics[width=\columnwidth]{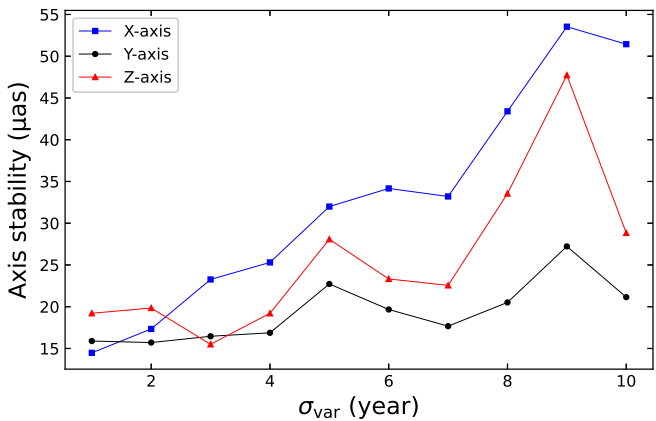}
        \caption{\label{fig:spin-vs-sigma}   
         Axis stability as a function of the amplitude of simulated positional drift.
         }
    \end{figure}

    We averaged the source positions weighted by their uncertainties from the coordinate time series within a one-year moving window with a step of one-year (i.e., without overlaps) since 1979.
    An underlying assumption (that may not always valid) is that the typical time scale for a source to show a significant displacement is one year.
    The averaged source positions of the ICRF3 defining sources within the same year window thus formed a yearly representation of the ICRF3.
    We only considered the yearly celestial frames with the number of the ICRF3 defining sources to be greater than six, which ruled out most yearly frames before 1986.
    By doing so, we formed 36 yearly celestial reference frames.
    
    Figure~\ref{fig:rot-yearly-ts} depicts the orientation offset angles of the yearly reference frame with respect to the ICRF3 $S/X$-band frame.
    The variation in the orientation offsets reduces significantly after 1995 and has further improved since 2002.
    The bumps during 2001--2002 in the $\epsilon_x$ are due to the unreliable position derived from few observations for some individual sources (e.g., 1004-500).
    The weighted mean values of the orientation offset around the X-, Y-, and Z- axes are $\mathrm{-4\,\mu as}$, $\mathrm{+28\,\mu as}$, and $\mathrm{+27\,\mu as}$, respectively, while the corresponding WRMS values with respect to the mean are approximately $\mathrm{13\,\mu as}$, $\mathrm{16\,\mu as}$, and $\mathrm{17\,\mu as}$.
    When only considering the data points after 1995, the WRMS is reduced to $\mathrm{11\,\mu as}$, $\mathrm{13\,\mu as}$, and $\mathrm{7\,\mu as}$ for the X-, Y-, and Z- axes, respectively, while the corresponding mean values are changed to $\mathrm{-3\,\mu as}$, $\mathrm{+11\,\mu as}$, and $\mathrm{-4\,\mu as}$.
    Looking at data after 2018, we found that the scatters of the orientation for three ICRF3 axes are further reduced to $\mathrm{3\,\mu as}$, $\mathrm{11\,\mu as}$, and $\mathrm{6\,\mu as}$, respectively.

    Another method of constructing the yearly representation of the ICRF3 is to run a series of yearly VLBI global solutions with identical configurations but truncate the data collecting window (starting from 1979) to a certain year.
    The positions of the ICRF3 defining sources in each global solution provide another yearly representation of the ICRF3.
    We tested this method and computed the deviation of the yearly celestial frame axes, which was fully consistent with what is presented in Fig.~\ref{fig:rot-yearly-ts} within the formal error.
    
    In short, the deviation of the yearly celestial reference frame axis is generally at the level of $\mathrm{10\,\mu as-20\,\mu as}$.
    We did not detect significant degradation in the stability of the ICRF3 axes since 2018.
    
\section{Discussion}

    As shown in the Eq.~(\ref{eq:spin-from-def-from-bs}) and also Fig.~\ref{fig:rot-yearly-ts}, there is no statistically significant rotation in the ICRF3, which is very satisfactory.
    One question risen is how much the axis stability would be changed by a very unstable defining source.
    To answer this question, we simulated the behavior of the emitting center due to the variability of the extragalatic source by a process with both Gaussian and Markovian characteristics as done in \citet{2016A&A...589A..71B}.
    The position drift at time $t_i$ was modeled as 
    %
    \begin{equation} \label{eq:gaussian-markovian}
        \left[\begin{array}{c}
        \Delta \alpha_{*}\left(t_{i}\right) \\
        \Delta \delta\left(t_{i}\right)
        \end{array}\right]=\mathrm{e}^{-\Delta t_{i} / \tau_{\mathrm{cor}}}\left[\begin{array}{c}
        \Delta \alpha_{*}\left(t_{i-1}\right) \\
        \Delta \delta\left(t_{i-1}\right)
        \end{array}\right]+\left[\begin{array}{c}
        g_{i}^{\alpha_{*}} \\
        g_{i}^{\delta}
        \end{array}\right],
    \end{equation}
    where $\Delta t_i=t_i-t_{i-1}$ is the step length, 
    $\tau_{\mathrm{cor}}$ is the characteristic correlation timescale, and
    $g_{i}^{\alpha_{*}}$ and $g_{i}^{\delta}$ are the Gaussian variables with zero mean and standard deviation 
    %
    \begin{equation} \label{eq:sigma}
    \sigma_{i}=\sigma_{\mathrm{var}} \sqrt{1-\exp \left(-2 \Delta t_{i} / \tau_{\mathrm{cor}}\right)}.
    \end{equation}
    The amplitude of the positional variation $\sigma_{\mathrm{var}}$ ranged from 1\,mas to 10\,mas in the simulation.
    We considered source 0552+398 as the test source since it has the largest number of observations.
    We assumed that $\tau_{\mathrm{cor}} = 5$ years.
    An example of the simulated position drift is given in Fig.~\ref{fig:0552+398-TS} in Appendix~\ref{sect:simulation-RW}.
    We considered the WRMS of the relative orientation of the yearly celestial frames (Sect.~\ref{subsec:orient-from-yearly-crf}) as the indicator of the axis stability.
    Figure~\ref{fig:spin-vs-sigma} depicts the relationship between these two quantities, from which we found a general increasing trend as expected.
    When $\sigma_{\mathrm{var}}\ge3\,$mas, the scatter of the X-axis orientation exceeds $\mathrm{20\,\mu as}$, which means that the frame becomes more unstable than the actual level of 10--20$\mathrm{\,\mu as\,yr^{-1}}$.
    Figure~\ref{fig:rot-yearly-ts-rw} presents the relative orientation of yearly celestial frames with respect to the ICRF3 when $\sigma_{\mathrm{var}}=3$\,mas.
    This simple simulation suggests a lower limit of 3\,mas for the amplitude of source position variation when the orientation angle of the yearly frames may yield a detectable instability.

\section{Conclusion}

    We evaluate the ICRF3 axis stability based on the coordinate time series of the extragalactic sources, which are derived from the geodetic VLBI observations from 1979--2020.
    The main results are remarked below.
    \begin{enumerate}
      \item The global spin inferred from the apparent proper motion of the ICRF3 defining sources is no greater than 0.8$\mathrm{\,\mu as\,yr^{-1}}$ for each axis with a formal error of $\mathrm{0.3\,\mu as\,yr^{-1}}$, which corresponds to a directional deformation of less than 30$\mathrm{\,\mu as}$ for the ICRF3 axes accumulated in 1979--2020.
      \item The scatter of the axis orientation for the yearly representation of ICRF3 is on the order of 10$\mathrm{\,\mu as}$--20$\mathrm{\,\mu as}$.
      There was no obvious degradation of stability for the ICRF3 axes after 2018 (i.e., the adoption of the ICRF3 as the fundamental reference frame).
    \end{enumerate}
    Therefore, the ICRF3 axes are found to be stable at the level of 10$\mathrm{\,\mu as}$--20$\mathrm{\,\mu as}$ during August 1979 and December 2020.
    
    In this work, the evaluation of the celestial frame axes stability is based on the source coordinate time series.
    This method captures both the mean variation of the axis orientation of the celestial frame and its time-dependent feature so that one can easily find and report the degradation of the axis stability if it occurs.
    In addition, this method is easy to launch and can regularly assess the axes stability, making it a good alternative method for monitoring the celestial frame axes.
    Considering that the variability exists for most extragalactic sources, we recommend regular assessments of the axes stability using the method based on the source coordinate time series.

\begin{acknowledgements}
    We sincerely thank Fran{\c c}ois Mignard for the constructive comments and useful suggestions, which improve the work a lot.
    N.L. and Z.Z. are supported by the National Natural Science Foundation of China (NSFC) under grant No. 11833004.
    N.L. is also supported by the Fundamental Research Funds for the Central Universities of China (grant No. 14380042) and China Postdoctoral Science Foundation (grant No. 2021M691530).
    This research had also made use of IPython \citep{2007CSE.....9c..21P}, Astropy\footnote{\href{http://www.astropy.org}{http://www.astropy.org}} \citep{2018AJ....156..123A}, the Python 2D plotting library Matplotlib \citep{2007CSE.....9...90H}, and NASA's Astrophysics Data System.
%
\end{acknowledgements}

\bibliographystyle{aa} 
\bibliography{references} 

\begin{thebibliography}{37}
\expandafter\ifx\csname natexlab\endcsname\relax\def\natexlab#1{#1}\fi

\bibitem[{{Andrei} {et~al.}(2009){Andrei}, {Bouquillon}, {de Camargo}, {Penna},
  {Taris}, {Souchay}, {da Silva Neto}, {Vieira Martins}, \&
  {Assafin}}]{2009jsrs.conf..199A}
{Andrei}, A.~H., {Bouquillon}, S., {de Camargo}, J.~I.~B., {et~al.} 2009, in
  Journ\'ees Syst\'emes de R\'ef\'erence Spatio-temporels 2008, ed. M.~{Soffel}
  \& N.~{Capitaine}, 199--202

\bibitem[{{Arias} \& {Bouquillon}(2004)}]{2004A&A...422.1105A}
{Arias}, E.~F. \& {Bouquillon}, S. 2004, \aap, 422, 1105

\bibitem[{{Astropy Collaboration} {et~al.}(2018){Astropy Collaboration},
  {Price-Whelan}, {Sip{\H{o}}cz}, {G{\"u}nther}, {Lim}, {Crawford}, {Conseil},
  {Shupe}, {Craig}, {Dencheva}, {Ginsburg}, {VanderPlas}, {Bradley},
  {P{\'e}rez-Su{\'a}rez}, {de Val-Borro}, {Aldcroft}, {Cruz}, {Robitaille},
  {Tollerud}, {Ardelean}, {Babej}, {Bach}, {Bachetti}, {Bakanov}, {Bamford},
  {Barentsen}, {Barmby}, {Baumbach}, {Berry}, {Biscani}, {Boquien}, {Bostroem},
  {Bouma}, {Brammer}, {Bray}, {Breytenbach}, {Buddelmeijer}, {Burke},
  {Calderone}, {Cano Rodr{\'\i}guez}, {Cara}, {Cardoso}, {Cheedella}, {Copin},
  {Corrales}, {Crichton}, {D'Avella}, {Deil}, {Depagne}, {Dietrich}, {Donath},
  {Droettboom}, {Earl}, {Erben}, {Fabbro}, {Ferreira}, {Finethy}, {Fox},
  {Garrison}, {Gibbons}, {Goldstein}, {Gommers}, {Greco}, {Greenfield},
  {Groener}, {Grollier}, {Hagen}, {Hirst}, {Homeier}, {Horton}, {Hosseinzadeh},
  {Hu}, {Hunkeler}, {Ivezi{\'c}}, {Jain}, {Jenness}, {Kanarek}, {Kendrew},
  {Kern}, {Kerzendorf}, {Khvalko}, {King}, {Kirkby}, {Kulkarni}, {Kumar},
  {Lee}, {Lenz}, {Littlefair}, {Ma}, {Macleod}, {Mastropietro}, {McCully},
  {Montagnac}, {Morris}, {Mueller}, {Mumford}, {Muna}, {Murphy}, {Nelson},
  {Nguyen}, {Ninan}, {N{\"o}the}, {Ogaz}, {Oh}, {Parejko}, {Parley}, {Pascual},
  {Patil}, {Patil}, {Plunkett}, {Prochaska}, {Rastogi}, {Reddy Janga},
  {Sabater}, {Sakurikar}, {Seifert}, {Sherbert}, {Sherwood-Taylor}, {Shih},
  {Sick}, {Silbiger}, {Singanamalla}, {Singer}, {Sladen}, {Sooley},
  {Sornarajah}, {Streicher}, {Teuben}, {Thomas}, {Tremblay}, {Turner},
  {Terr{\'o}n}, {van Kerkwijk}, {de la Vega}, {Watkins}, {Weaver}, {Whitmore},
  {Woillez}, {Zabalza}, \& {Astropy Contributors}}]{2018AJ....156..123A}
{Astropy Collaboration}, {Price-Whelan}, A.~M., {Sip{\H{o}}cz}, B.~M., {et~al.}
  2018, \aj, 156, 123

\bibitem[{{Bachchan} {et~al.}(2016){Bachchan}, {Hobbs}, \&
  {Lindegren}}]{2016A&A...589A..71B}
{Bachchan}, R.~K., {Hobbs}, D., \& {Lindegren}, L. 2016, \aap, 589, A71

\bibitem[{{Charlot} {et~al.}(2020){Charlot}, {Jacobs}, {Gordon}, {Lambert}, {de
  Witt}, {B{\"o}hm}, {Fey}, {Heinkelmann}, {Skurikhina}, {Titov}, {Arias},
  {Bolotin}, {Bourda}, {Ma}, {Malkin}, {Nothnagel}, {Mayer}, {MacMillan},
  {Nilsson}, \& {Gaume}}]{2020A&A...644A.159C}
{Charlot}, P., {Jacobs}, C.~S., {Gordon}, D., {et~al.} 2020, \aap, 644, A159

\bibitem[{{Dehant} {et~al.}(2003){Dehant}, {Feissel-Vernier}, {de Viron}, {Ma},
  {Yseboodt}, \& {Bizouard}}]{2003JGRB..108.2275D}
{Dehant}, V., {Feissel-Vernier}, M., {de Viron}, O., {et~al.} 2003, Journal of
  Geophysical Research (Solid Earth), 108, 2275

\bibitem[{{Feissel-Vernier}(2003)}]{2003A&A...403..105F}
{Feissel-Vernier}, M. 2003, \aap, 403, 105

\bibitem[{{Feissel-Vernier} {et~al.}(2006){Feissel-Vernier}, {Ma}, {Gontier},
  \& {Barache}}]{2006A&A...452.1107F}
{Feissel-Vernier}, M., {Ma}, C., {Gontier}, A.~M., \& {Barache}, C. 2006, \aap,
  452, 1107

\bibitem[{{Fey} {et~al.}(2015){Fey}, {Gordon}, {Jacobs}, {Ma}, {Gaume},
  {Arias}, {Bianco}, {Boboltz}, {B{\"o}ckmann}, {Bolotin}, {Charlot},
  {Collioud}, {Engelhardt}, {Gipson}, {Gontier}, {Heinkelmann}, {Kurdubov},
  {Lambert}, {Lytvyn}, {MacMillan}, {Malkin}, {Nothnagel}, {Ojha},
  {Skurikhina}, {Sokolova}, {Souchay}, {Sovers}, {Tesmer}, {Titov}, {Wang}, \&
  {Zharov}}]{2015AJ....150...58F}
{Fey}, A.~L., {Gordon}, D., {Jacobs}, C.~S., {et~al.} 2015, \aj, 150, 58

\bibitem[{{Fomalont} {et~al.}(2011){Fomalont}, {Johnston}, {Fey}, {Boboltz},
  {Oyama}, \& {Honma}}]{2011AJ....141...91F}
{Fomalont}, E., {Johnston}, K., {Fey}, A., {et~al.} 2011, \aj, 141, 91

\bibitem[{{Gaia Collaboration} {et~al.}(2018){Gaia Collaboration}, {Mignard},
  {Klioner}, {Lindegren}, {Hern{\'a}ndez}, {Bastian}, {Bombrun}, {Hobbs},
  {Lammers}, {Michalik}, {Ramos-Lerate}, {Biermann},
  {Fern{\'a}ndez-Hern{\'a}ndez}, {Geyer}, {Hilger}, {Siddiqui},
  {Steidelm{\"u}ller}, {Babusiaux}, {Barache}, {Lambert}, {Andrei}, {Bourda},
  {Charlot}, {Brown}, {Vallenari}, {Prusti}, {de Bruijne}, {Bailer-Jones},
  {Evans}, {Eyer}, {Jansen}, {Jordi}, {Luri}, {Panem}, {Pourbaix}, {Randich},
  {Sartoretti}, {Soubiran}, {van Leeuwen}, {Walton}, {Arenou}, {Cropper},
  {Drimmel}, {Katz}, {Lattanzi}, {Bakker}, {Cacciari}, {Casta{\~n}eda},
  {Chaoul}, {Cheek}, {De Angeli}, {Fabricius}, {Guerra}, {Holl}, {Masana},
  {Messineo}, {Mowlavi}, {Nienartowicz}, {Panuzzo}, {Portell}, {Riello},
  {Seabroke}, {Tanga}, {Th{\'e}venin}, {Gracia-Abril}, {Comoretto},
  {Garcia-Reinaldos}, {Teyssier}, {Altmann}, {Andrae}, {Audard},
  {Bellas-Velidis}, {Benson}, {Berthier}, {Blomme}, {Burgess}, {Busso},
  {Carry}, {Cellino}, {Clementini}, {Clotet}, {Creevey}, {Davidson}, {De
  Ridder}, {Delchambre}, {Dell'Oro}, {Ducourant}, {Fouesneau}, {Fr{\'e}mat},
  {Galluccio}, {Garc{\'\i}a-Torres}, {Gonz{\'a}lez-N{\'u}{\~n}ez},
  {Gonz{\'a}lez-Vidal}, {Gosset}, {Guy}, {Halbwachs}, {Hambly}, {Harrison},
  {Hestroffer}, {Hodgkin}, {Hutton}, {Jasniewicz}, {Jean-Antoine-Piccolo},
  {Jordan}, {Korn}, {Krone-Martins}, {Lanzafame}, {Lebzelter}, {L{\"o}ffler},
  {Manteiga}, {Marrese}, {Mart{\'\i}n-Fleitas}, {Moitinho}, {Mora}, {Muinonen},
  {Osinde}, {Pancino}, {Pauwels}, {Petit}, {Recio-Blanco}, {Richards},
  {Rimoldini}, {Robin}, {Sarro}, {Siopis}, {Smith}, {Sozzetti}, {S{\"u}veges},
  {Torra}, {van Reeven}, {Abbas}, {Abreu Aramburu}, {Accart}, {Aerts},
  {Altavilla}, {{\'A}lvarez}, {Alvarez}, {Alves}, {Anderson}, {Anglada Varela},
  {Antiche}, {Antoja}, {Arcay}, {Astraatmadja}, {Bach}, {Baker},
  {Balaguer-N{\'u}{\~n}ez}, {Balm}, {Barata}, {Barbato}, {Barblan}, {Barklem},
  {Barrado}, {Barros}, {Barstow}, {Bartholom{\'e} Mu{\~n}oz}, {Bassilana},
  {Becciani}, {Bellazzini}, {Berihuete}, {Bertone}, {Bianchi}, {Bienaym{\'e}},
  {Blanco-Cuaresma}, {Boch}, {Boeche}, {Borrachero}, {Bossini}, {Bouquillon},
  {Bragaglia}, {Bramante}, {Breddels}, {Bressan}, {Brouillet},
  {Br{\"u}semeister}, {Brugaletta}, {Bucciarelli}, {Burlacu}, {Busonero},
  {Butkevich}, {Buzzi}, {Caffau}, {Cancelliere}, {Cannizzaro}, {Cantat-Gaudin},
  {Carballo}, {Carlucci}, {Carrasco}, {Casamiquela}, {Castellani},
  {Castro-Ginard}, {Chemin}, {Chiavassa}, {Cocozza}, {Costigan}, {Cowell},
  {Crifo}, {Crosta}, {Crowley}, {Cuypers}, {Dafonte}, {Damerdji}, {Dapergolas},
  {David}, {David}, {de Laverny}, {De Luise}, {De March}, {de Souza}, {de
  Torres}, {Debosscher}, {del Pozo}, {Delbo}, {Delgado}, {Delgado}, {Diakite},
  {Diener}, {Distefano}, {Dolding}, {Drazinos}, {Dur{\'a}n}, {Edvardsson},
  {Enke}, {Eriksson}, {Esquej}, {Eynard Bontemps}, {Fabre}, {Fabrizio},
  {Faigler}, {Falc{\~a}o}, {Farr{\`a}s Casas}, {Federici}, {Fedorets},
  {Fernique}, {Figueras}, {Filippi}, {Findeisen}, {Fonti}, {Fraile}, {Fraser},
  {Fr{\'e}zouls}, {Gai}, {Galleti}, {Garabato}, {Garc{\'\i}a-Sedano},
  {Garofalo}, {Garralda}, {Gavel}, {Gavras}, {Gerssen}, {Giacobbe}, {Gilmore},
  {Girona}, {Giuffrida}, {Glass}, {Gomes}, {Granvik}, {Gueguen}, {Guerrier},
  {Guiraud}, {Guti{\'e}}, {Haigron}, {Hatzidimitriou}, {Hauser}, {Haywood},
  {Heiter}, {Helmi}, {Heu}, {Hofmann}, {Holland}, {Huckle}, {Hypki}, {Icardi},
  {Jan{\ss}en}, {Jevardat de Fombelle}, {Jonker}, {Juh{\'a}sz}, {Julbe},
  {Karampelas}, {Kewley}, {Klar}, {Kochoska}, {Kohley}, {Kolenberg},
  {Kontizas}, {Kontizas}, {Koposov}, {Kordopatis}, {Kostrzewa-Rutkowska},
  {Koubsky}, {Lanza}, {Lasne}, {Lavigne}, {Le Fustec}, {Le Poncin-Lafitte},
  {Lebreton}, {Leccia}, {Leclerc}, {Lecoeur-Taibi}, {Lenhardt}, {Leroux},
  {Liao}, {Licata}, {Lindstr{\o}m}, {Lister}, {Livanou}, {Lobel}, {L{\'o}pez},
  {Managau}, {Mann}, {Mantelet}, {Marchal}, {Marchant}, {Marconi}, {Marinoni},
  {Marschalk{\'o}}, {Marshall}, {Martino}, {Marton}, {Mary}, {Massari},
  {Matijevi{\v{c}}}, {Mazeh}, {McMillan}, {Messina}, {Millar}, {Molina},
  {Molinaro}, {Moln{\'a}r}, {Montegriffo}, {Mor}, {Morbidelli}, {Morel},
  {Morris}, {Mulone}, {Muraveva}, {Musella}, {Nelemans}, {Nicastro}, {Noval},
  {O'Mullane}, {Ord{\'e}novic}, {Ord{\'o}{\~n}ez-Blanco}, {Osborne}, {Pagani},
  {Pagano}, {Pailler}, {Palacin}, {Palaversa}, {Panahi}, {Pawlak},
  {Piersimoni}, {Pineau}, {Plachy}, {Plum}, {Poggio}, {Poujoulet},
  {Pr{\v{s}}a}, {Pulone}, {Racero}, {Ragaini}, {Rambaux}, {Regibo},
  {Reyl{\'e}}, {Riclet}, {Ripepi}, {Riva}, {Rivard}, {Rixon}, {Roegiers},
  {Roelens}, {Romero-G{\'o}mez}, {Rowell}, {Royer}, {Ruiz-Dern}, {Sadowski},
  {Sagrist{\`a} Sell{\'e}s}, {Sahlmann}, {Salgado}, {Salguero}, {Sanna},
  {Santana-Ros}, {Sarasso}, {Savietto}, {Schultheis}, {Sciacca}, {Segol},
  {Segovia}, {S{\'e}gransan}, {Shih}, {Siltala}, {Silva}, {Smart}, {Smith},
  {Solano}, {Solitro}, {Sordo}, {Soria Nieto}, {Souchay}, {Spagna}, {Spoto},
  {Stampa}, {Steele}, {Stephenson}, {Stoev}, {Suess}, {Surdej}, {Szabados},
  {Szegedi-Elek}, {Tapiador}, {Taris}, {Tauran}, {Taylor}, {Teixeira},
  {Terrett}, {Teyssandier}, {Thuillot}, {Titarenko}, {Torra Clotet}, {Turon},
  {Ulla}, {Utrilla}, {Uzzi}, {Vaillant}, {Valentini}, {Valette}, {van Elteren},
  {Van Hemelryck}, {van Leeuwen}, {Vaschetto}, {Vecchiato}, {Veljanoski},
  {Viala}, {Vicente}, {Vogt}, {von Essen}, {Voss}, {Votruba}, {Voutsinas},
  {Walmsley}, {Weiler}, {Wertz}, {Wevers}, {Wyrzykowski}, {Yoldas},
  {{\v{Z}}erjal}, {Ziaeepour}, {Zorec}, {Zschocke}, {Zucker}, {Zurbach}, \&
  {Zwitter}}]{2018A&A...616A..14G}
{Gaia Collaboration}, {Mignard}, F., {Klioner}, S.~A., {et~al.} 2018, \aap,
  616, A14

\bibitem[{{Gattano} \& {Charlot}(2021)}]{2021A&A...648A.125G}
{Gattano}, C. \& {Charlot}, P. 2021, \aap, 648, A125

\bibitem[{{Gattano} {et~al.}(2018){Gattano}, {Lambert}, \& {Le
  Bail}}]{2018A&A...618A..80G}
{Gattano}, C., {Lambert}, S.~B., \& {Le Bail}, K. 2018, \aap, 618, A80

\bibitem[{{Hunter}(2007)}]{2007CSE.....9...90H}
{Hunter}, J.~D. 2007, Computing in Science and Engineering, 9, 90

\bibitem[{{Lambert}(2013)}]{2013A&A...553A.122L}
{Lambert}, S. 2013, \aap, 553, A122

\bibitem[{{Lambert}(2014)}]{2014A&A...570A.108L}
{Lambert}, S. 2014, \aap, 570, A108

\bibitem[{{Lambert} {et~al.}(2008){Lambert}, {Dehant}, \&
  {Gontier}}]{2008A&A...481..535L}
{Lambert}, S.~B., {Dehant}, V., \& {Gontier}, A.~M. 2008, \aap, 481, 535

\bibitem[{{Lambert} \& {Gontier}(2009)}]{2009A&A...493..317L}
{Lambert}, S.~B. \& {Gontier}, A.~M. 2009, \aap, 493, 317

\bibitem[{{Larchenkova} {et~al.}(2017){Larchenkova}, {Lutovinov}, \&
  {Lyskova}}]{2017ApJ...835...51L}
{Larchenkova}, T.~I., {Lutovinov}, A.~A., \& {Lyskova}, N.~S. 2017, \apj, 835,
  51

\bibitem[{{Larchenkova} {et~al.}(2020){Larchenkova}, {Lyskova}, {Petrov}, \&
  {Lutovinov}}]{2020ApJ...898...51L}
{Larchenkova}, T.~I., {Lyskova}, N.~S., {Petrov}, L., \& {Lutovinov}, A.~A.
  2020, \apj, 898, 51

\bibitem[{{Liu} {et~al.}(2018){Liu}, {Malkin}, \& {Zhu}}]{2018MNRAS.474.4477L}
{Liu}, J.~C., {Malkin}, Z., \& {Zhu}, Z. 2018, \mnras, 474, 4477

\bibitem[{{Liu} {et~al.}(2017){Liu}, {Liu}, \& {Zhu}}]{2017MNRAS.466.1567L}
{Liu}, N., {Liu}, J.~C., \& {Zhu}, Z. 2017, \mnras, 466, 1567

\bibitem[{{Ma} {et~al.}(1998){Ma}, {Arias}, {Eubanks}, {Fey}, {Gontier},
  {Jacobs}, {Sovers}, {Archinal}, \& {Charlot}}]{1998AJ....116..516M}
{Ma}, C., {Arias}, E.~F., {Eubanks}, T.~M., {et~al.} 1998, \aj, 116, 516

\bibitem[{{Ma} {et~al.}(1986){Ma}, {Clark}, {Ryan}, {Herring}, {Shapiro},
  {Corey}, {Hinteregger}, {Rogers}, {Whitney}, {Knight}, {Lundqvist},
  {Shaffer}, {Vandenberg}, {Pigg}, {Schupler}, \&
  {Ronnang}}]{1986AJ.....92.1020M}
{Ma}, C., {Clark}, T.~A., {Ryan}, J.~W., {et~al.} 1986, \aj, 92, 1020

\bibitem[{{MacMillan} {et~al.}(2019){MacMillan}, {Fey}, {Gipson}, {Gordon},
  {Jacobs}, {Kr{\'a}sn{\'a}}, {Lambert}, {Malkin}, {Titov}, {Wang}, \&
  {Xu}}]{2019A&A...630A..93M}
{MacMillan}, D.~S., {Fey}, A., {Gipson}, J.~M., {et~al.} 2019, \aap, 630, A93

\bibitem[{{Mignard} \& {Klioner}(2012)}]{2012A&A...547A..59M}
{Mignard}, F. \& {Klioner}, S. 2012, \aap, 547, A59

\bibitem[{{Mignard} {et~al.}(2016){Mignard}, {Klioner}, {Lindegren}, {Bastian},
  {Bombrun}, {Hern{\'a}ndez}, {Hobbs}, {Lammers}, {Michalik}, {Ramos-Lerate},
  {Biermann}, {Butkevich}, {Comoretto}, {Joliet}, {Holl}, {Hutton}, {Parsons},
  {Steidelm{\"u}ller}, {Andrei}, {Bourda}, \& {Charlot}}]{2016A&A...595A...5M}
{Mignard}, F., {Klioner}, S., {Lindegren}, L., {et~al.} 2016, \aap, 595, A5

\bibitem[{{Mo{\'o}r} {et~al.}(2011){Mo{\'o}r}, {Frey}, {Lambert}, {Titov}, \&
  {Bakos}}]{2011AJ....141..178M}
{Mo{\'o}r}, A., {Frey}, S., {Lambert}, S.~B., {Titov}, O.~A., \& {Bakos}, J.
  2011, \aj, 141, 178

\bibitem[{{Nothnagel} {et~al.}(2017){Nothnagel}, {Artz}, {Behrend}, \&
  {Malkin}}]{2017JGeod..91..711N}
{Nothnagel}, A., {Artz}, T., {Behrend}, D., \& {Malkin}, Z. 2017, Journal of
  Geodesy, 91, 711

\bibitem[{{Perez} \& {Granger}(2007)}]{2007CSE.....9c..21P}
{Perez}, F. \& {Granger}, B.~E. 2007, Computing in Science and Engineering, 9,
  21

\bibitem[{{Plank} {et~al.}(2016){Plank}, {Shabala}, {McCallum},
  {Kr{\'a}sn{\'a}}, {Petrachenko}, {Rastorgueva-Foi}, \&
  {Lovell}}]{2016MNRAS.455..343P}
{Plank}, L., {Shabala}, S.~S., {McCallum}, J.~N., {et~al.} 2016, \mnras, 455,
  343

\bibitem[{{Sazhin} {et~al.}(1998){Sazhin}, {Zharov}, {Volynkin}, \&
  {Kalinina}}]{1998MNRAS.300..287S}
{Sazhin}, M.~V., {Zharov}, V.~E., {Volynkin}, A.~V., \& {Kalinina}, T.~A. 1998,
  \mnras, 300, 287

\bibitem[{{Schaap} {et~al.}(2013){Schaap}, {Shabala}, {Ellingsen}, {Titov}, \&
  {Lovell}}]{2013MNRAS.434..585S}
{Schaap}, R.~G., {Shabala}, S.~S., {Ellingsen}, S.~P., {Titov}, O.~A., \&
  {Lovell}, J.~E.~J. 2013, \mnras, 434, 585

\bibitem[{{Shabala} {et~al.}(2014){Shabala}, {Rogers}, {McCallum}, {Titov},
  {Blanchard}, {Lovell}, \& {Watson}}]{2014JGeod..88..575S}
{Shabala}, S.~S., {Rogers}, J.~G., {McCallum}, J.~N., {et~al.} 2014, Journal of
  Geodesy, 88, 575

\bibitem[{{Taris} {et~al.}(2013){Taris}, {Andrei}, {Klotz}, {Vachier},
  {C{\^o}te}, {Bouquillon}, {Souchay}, {Lambert}, {Anton}, {Bourda}, \&
  {Coward}}]{2013A&A...552A..98T}
{Taris}, F., {Andrei}, A., {Klotz}, A., {et~al.} 2013, \aap, 552, A98

\bibitem[{{Taris} {et~al.}(2016){Taris}, {Andrei}, {Roland}, {Klotz},
  {Vachier}, \& {Souchay}}]{2016A&A...587A.112T}
{Taris}, F., {Andrei}, A., {Roland}, J., {et~al.} 2016, \aap, 587, A112

\bibitem[{{Taris} {et~al.}(2018){Taris}, {Damljanovic}, {Andrei}, {Souchay},
  {Klotz}, \& {Vachier}}]{2018A&A...611A..52T}
{Taris}, F., {Damljanovic}, G., {Andrei}, A., {et~al.} 2018, \aap, 611, A52

\end{thebibliography}

\begin{appendix} 

\section{Simulation of positional variation for source 0552+398} \label{sect:simulation-RW}

 \begin{figure}[hbtp]
  \centering
  \includegraphics[width=\columnwidth]{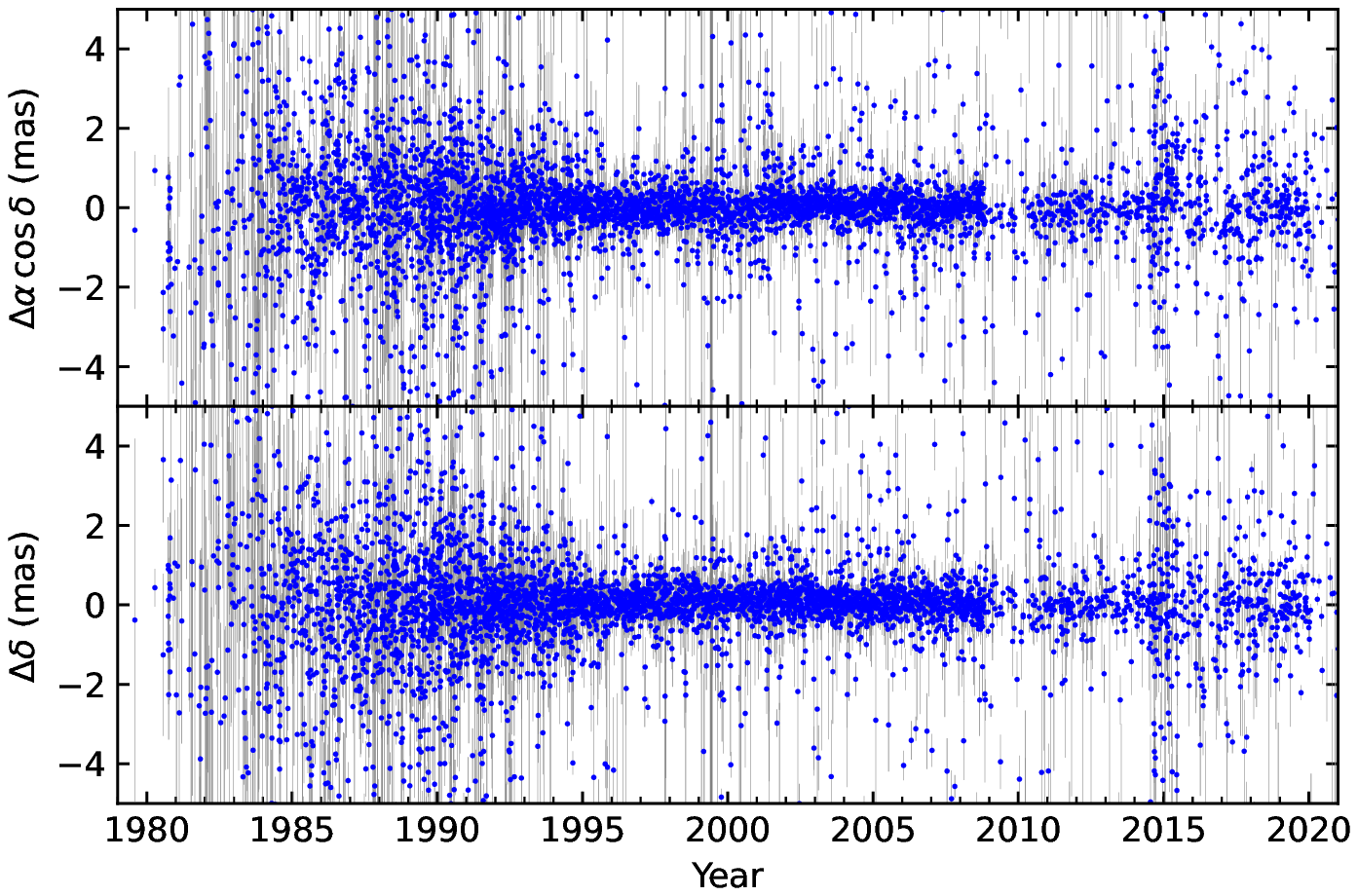}  
  \includegraphics[width=\columnwidth]{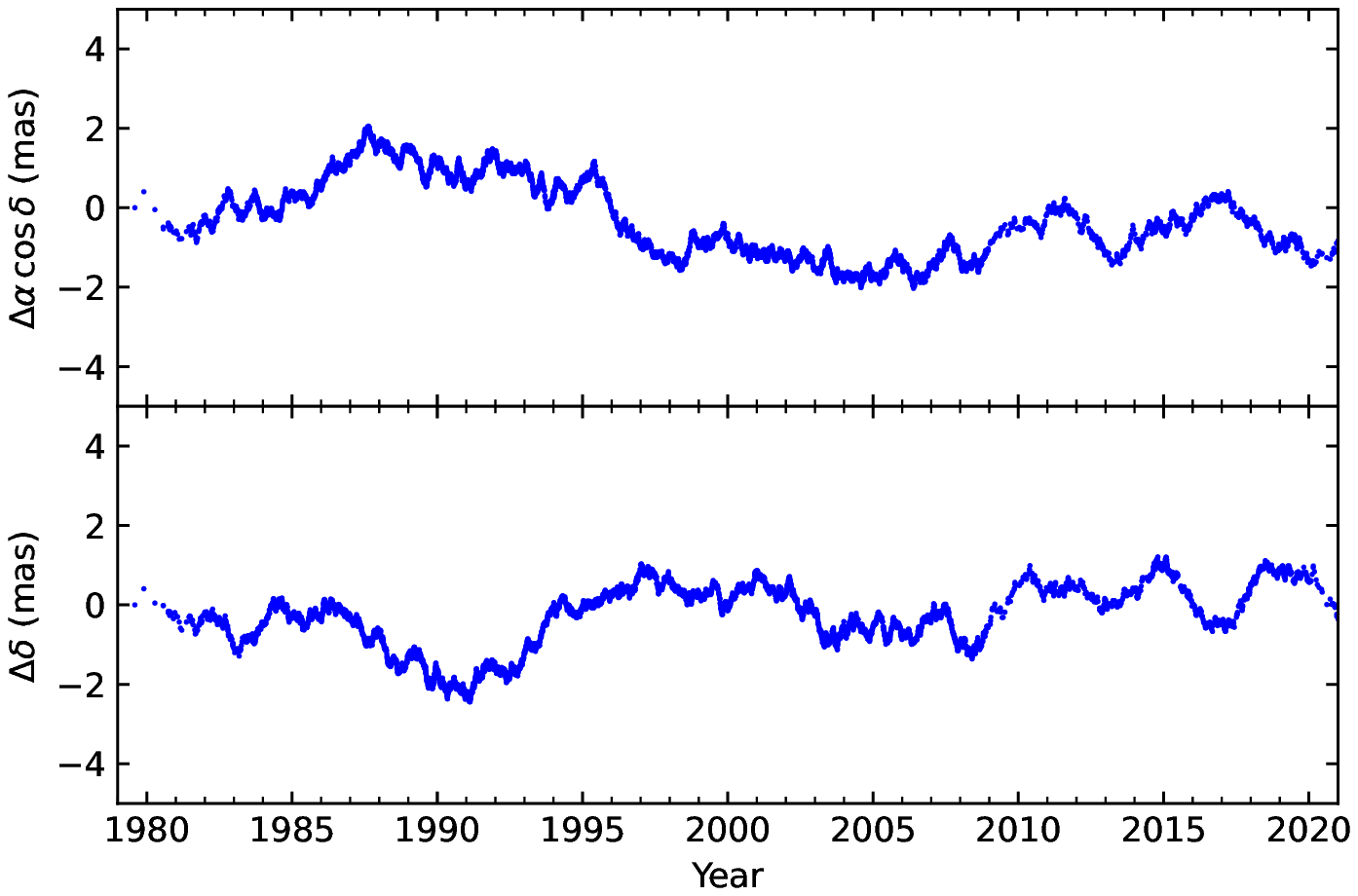}  
  \caption[]{\label{fig:0552+398-TS} %
  Top: the coordinate offset time series of source 0552+398 with referred to its position in the ICRF3 $S/X$-band catalog.
  Bottom: the simulated position drift due to the photometric variability.
  We assumed that $\tau_{\mathrm{cor}}=5$ years and $\sigma_{\rm var}=3$\,mas. 
  }
 \end{figure}
    
    \begin{figure}
        \centering
        \includegraphics[width=\columnwidth]{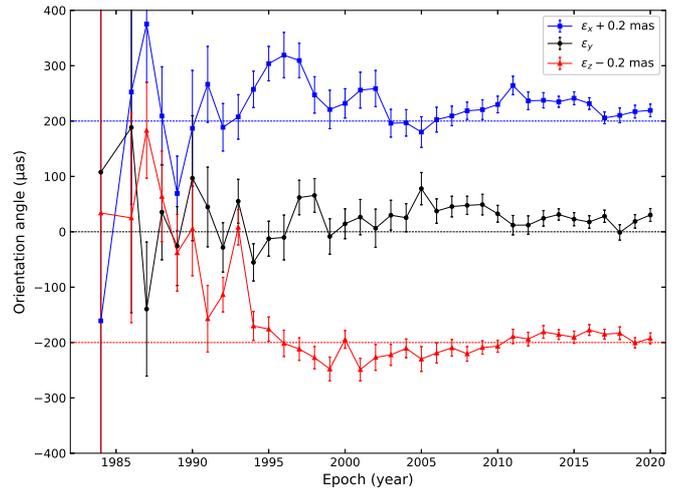}
        \caption{\label{fig:rot-yearly-ts-rw}   
         Relative orientation of yearly celestial reference frames with respect to the ICRF3 $S/X$-band frame in the case that source 0552+398 shows a position drift with an amplitude of 3\,mas. 
         }
    \end{figure}

    Figure~\ref{fig:0552+398-TS} presents the coordinate time series of 0552+398 derived from the VLBI observations (top) and also simulated using Eq.~\ref{eq:gaussian-markovian} (bottom).
    If the simulated time series replace the real ones, the stability of the axes orientation of the yearly celestial frames would become worse, as shown in Fig.~\ref{fig:rot-yearly-ts-rw}.
    
\end{appendix}

\end{document}